\documentclass[%
 reprint,
 pre,
 amsmath,amssymb,
]{revtex4-2}

\usepackage{graphicx}
\usepackage{dcolumn}
\usepackage{bm}
\usepackage[dvipsnames]{xcolor}
\usepackage[colorlinks]{hyperref}
\hypersetup{colorlinks=true,allcolors=Blue}
\usepackage{soul}

\usepackage{xcolor}

\newcommand{\vect}{\boldsymbol}
\newcommand{\mobD}{\Lambda}

\newcommand{\figref}[1]{Fig.~\ref{#1}}
\newcommand{\Eqref}[1]{Eq.~\eqref{#1}}
\newcommand{\Eqsref}[1]{Eqs.~\eqref{#1}}
\newcommand{\cIn}{c_\mathrm{in}}
\newcommand{\cInN}{c_\mathrm{in}^{(0)}}
\newcommand{\cOut}{c_\mathrm{out}}
\newcommand{\cOutN}{c_\mathrm{out}^{(0)}}

\newcommand{\cSub}{c_\mathrm{sub}}

\newcommand{\Vsub}{V_\mathrm{sub}}

\newcommand{\kBT}{k_\mathrm{B} T}

\newcommand{\mrm}[1]{\mathrm{#1}}
\newcommand{\diff}{\mathrm{d}}
\newcommand{\Vsys}{V_\mathrm{sys}}

\begin{document}
\preprint{APS/123-QED}

\title{Chemically Active Liquid Bridges Generate Repulsive Forces}%

\author{Noah Ziethen}
\author{Frieder Johannsen}
\author{David Zwicker}%
 \email{david.zwicker@ds.mpg.de}
\affiliation{%
 Max Planck Institute for Dynamics and Self-Organization\\
 Am Faßberg 17 37077 Göttingen
}%

\date{\today}

\begin{abstract}
    Droplets help organize cells by compartmentalizing biomolecules and by mediating mechanical interactions. When bridging two structures, such droplets generate capillary forces, which depend on surface properties and distance. While the forces exerted by passive liquid bridges are well understood, the role of active chemical reactions, which are often present in biological droplets, remains unclear. To elucidate this case, we study a single liquid bridge with continuous chemical turnover. These reactions control the bridge radius and lead to purely repulsive forces—contrasting with the typically attractive forces in passive systems. Our results reveal how chemical activity can fundamentally alter forces generated by liquid bridges, which could be exploited by cells.
\end{abstract}

\maketitle

\tableofcontents

\section{Introduction}

Droplets forming by phase separation are crucial for organizing the inside of biological cells, e.g., by partitioning molecules, controlling reactions, and generating forces~\cite{Brangwynne_Eckmann_Courson_Rybarska_Hoege_Gharakhani_Julicher_Hyman_2009,Hyman2014,Banani_Lee_Hyman_Rosen_2017,Dignon2020,Su2021}.
Cells actively regulate phase separation through chemical reactions that modify the droplets' constituents~\cite{Hondele2019,Snead2019,Soeding_Zwicker_Sohrabi-Jahromi_Boehning_Kirschbaum_2019}.
Theoretical studies have shown that such chemical reactions can control droplet nucleation, size, and morphology~\cite{Zwicker_Decker_Jaensch_Hyman_Juelicher_2014,Zwicker_Suppression_2015,Zwicker_Seyboldt_Weber_Hyman_Juelicher_2017,Wurtz_Lee_2018_2,Weber_Zwicker_2019,Kirschbaum_Zwicker_2021,Zwicker2022a, Ziethen_Kirschbaum_Zwicker_2023, Ziethen_Zwicker_2024}.
Inside cells, such chemically active droplets interact with other structures, including the cytoskeleton and membranes
\cite{Gall_Bellini_Wu_Murphy_1999,Kaur_Raju_Alshareedah_Davis_Potoyan_Banerjee_2021,Setru_Gouveia_Alfaro-Aco_Shaevitz_Stone_Petry_2021,Kusumaatmaja_May_Feeney_McKenna_Mizushima_Frigerio_Knorr_2021,Quail_Golfier_Elsner_Ishihara_Murugesan_Renger_Juelicher_Brugues_2021}.
In particular, droplets can form capillary connections that generate mechanical forces between surfaces.
While the physics of passive liquid bridges is well understood, capillary forces resulting from active chemical reactions remain elusive.

In passive systems, the interactions between droplet and surface determines the macroscopic wetting angle~$\vartheta$, which gives a boundary condition to the minimization of the surface energy ultimately dictating the droplet's shape~\cite{fortes_axisymmetric_1982, Bormashenko_2009}.
In case of a liquid bridge, $\vartheta$ controls its shape, and the tendency to minimize surface energies generates a force, which is typically attractive~\cite{fortes_axisymmetric_1982, gouveia_capillary_2022}.
Such force generation was observed in recent experiments, where biomolecular condensates interacted with DNA~\cite{Zhao2025,Strom2024,Quail2021}.
Although non-equilibrium conditions were considered in Ref.~\cite{Strom2024}, the involved condensates did not exhibit any reactions, so the observed forces were still governed by passive mechanisms.

We ask how chemical activity affect forces generated by liquid bridges.
\section{Model}
\label{sec:model}
\begin{figure}
    \centering
    \includegraphics[width=\linewidth]{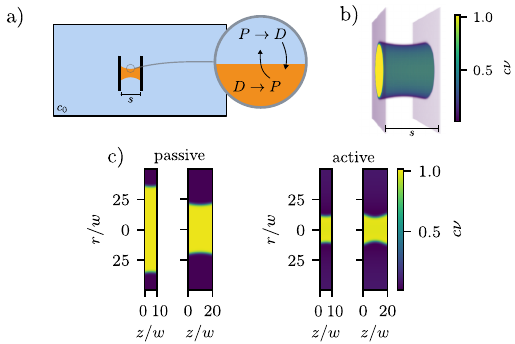}
    \caption{\textbf{Chemically active liquid bridges mediate interactions between walls.} \textrm{(a) 
            Schematic representation of the simulation setup comprising a subsystem with a liquid bridge (orange) bounded by two parallel planar walls embedded in a larger domain at concentration $c_0$. Active interconversion between precursor material $P$ and droplet material $D$ occurs throughout the system.
        (b) Cylindrically-symmetric density fields $c(\vect r)$ obtained from minimizing \Eqref{eq:tot_en_func}.
        (c) Slices of $c(\vect r)$ showing bridge configurations without (left) and with (right) chemical reactions for various wall separations $s$. Model parameters are $c_0=0.03 \nu^{-1}$, $a=\nu^3 \kBT$, $k_0=0$ for the passive case and $c_0=0.1\nu^{-1}$, $a=\nu^3 \kBT$, $k_0=0.001\mobD a \nu^{-2} w^{-2}$ for the active one.
}}
    \label{fig:setup}
\end{figure}

To do this, we consider a system of fixed volume $V_\mathrm{sys}$ filled with an incompressible, binary mixture of droplet and solvent material of equal molecular volume $\nu$ kept at constant temperature $T$. 
We embed into the system two movable walls interacting with the droplet material such that a liquid bridge can form between them via phase separation (\figref{fig:setup}).
We describe the phase separation of the droplet material density $c(\vect{r})$ using a free energy comprising bulk and surface terms~\cite{Cahn_Hilliard},
\begin{align}
    H=\int_{V_\mathrm{sys}} \left[h(c) + \frac{\kappa}{2}|\nabla c|^2\right]\mathrm{dV} + \int_{\partial V_\mathrm{sys}} g(c) \mathrm{dA}\;,
  \label{eq:free_en_surf}
\end{align}
where $h(c)$ is the local free energy density and $\kappa$ penalizes compositional gradients.
For simplicity, we consider
\begin{align}
    h(c) = \frac{a}{2}c^2\left(c-\frac{1}{\nu}\right)^2
	\label{eq:f_loc}
	\;,
\end{align}
where $a$ is an energy scale controlling the strength of phase separation.
We describe the interaction of the droplet material with the surface of the walls by a contact potential $g(c)$, which we for simplicity expand to linear order in $c$, $g(c) = g_0 + g_1c$~\cite{de_Gennes_1985}.

Minimizing the free energy given by \Eqsref{eq:free_en_surf}--\eqref{eq:f_loc} leads to two equilibrium conditions,
\begin{subequations}
\begin{align}
    h'(c) - \kappa \nabla^2 c &= \text{const} &&\text{(in the bulk),}\\
    \vect n \cdot \nabla c &=\frac{g_1}{\kappa} &&\text{(at the walls),}
  \label{eq:boundary_cond}
\end{align}
\end{subequations}
where $\vect n$ is the outward normal of the wall surface.
The first equation describes the balance of the exchange chemical potential $\mu = h'(c) - \kappa \nabla^2 c$, where the constant is determined from material conservation~\cite{Weber_Zwicker_2019}.
In thermodynamically large systems, this generically yields a dense phase of concentration $c^{(0)}_\mathrm{in}\nu = 1$ that is separated from a dilute phase of concentration $c^{(0)}_\mathrm{out}\nu = 0$ by an interface of width~$w=\nu\sqrt{\kappa /a}$~\cite{weber_physics_2019}.
In contrast, \Eqref{eq:boundary_cond} describes interactions of droplet and wall.
The Young-Dupr\'{e} equation relates this condition to the wetting angle $\vartheta$, $\gamma_\mathrm{ws}=\gamma_\mathrm{wd}+\gamma_\mathrm{ds}\cos(\vartheta)$, where $\gamma_{ij}$ denotes the pairwise surface tension between the wall, droplet, and solvent, respectively~\cite{Young1805}, implying $\cos(\vartheta) \approx 6 g_1\nu^2/\sqrt{\kappa a}$ \cite{Ziethen_Zwicker_2024}.
Since we want to analyze an isolated bridge, we focus on partial wetting ($0<\vartheta<\pi$), and consider a system inside the nucleation-and-growth regime, where the homogeneous state is metastable.

The final ingredient of our model is chemical activity, which describes the production and degradation of droplet material.
The dynamics of this active system are governed by the continuity equation,
\begin{align}
    \partial_t c + \nabla \cdot \vect{j} = k(c_0 - c) \; ,
    \label{eq:cahn-hilliard-source}
\end{align}
where $\vect j = - \mobD \nabla \mu$ accounts for diffusive fluxes with  mobility $\mobD$, whereas the right hand side describes  chemical reactions.
For simplicity, we  focus on linear reactions, which maintain an average composition $c_0$ at rate $k$.
Such reactions can be derived from thermodynamically-consistent models~\cite{Ziethen_Kirschbaum_Zwicker_2023} and are the stereotypical example for size-controlled droplets~\cite{Zwicker_Paulin_Burg_2025, Weber_Zwicker_2019, Zwicker_Suppression_2015}.

To investigate capillary forces in the active system, we map the dynamics given by \Eqref{eq:cahn-hilliard-source} to a passive system, $\partial_t c = \mobD \nabla^2 \delta \tilde{H}/\delta c$, introducing the augmented energy 
\begin{align}
       \tilde H[c] = H[c]  + \frac{k}{2 \mobD} \int \bigl[c(\vect r) - c_0\bigr]\Psi(\vect r) \, \mathrm{d}\vect{r}
       \;,
       \label{eq:tot_en_func}
\end{align}
where $\Psi$ solves the Poisson equation $\nabla^2 \Psi = c_0-c(\vect r)$ with Neumann boundary conditions ($\vect{n}.\nabla \Psi=0$)~\cite{Liu_Goldenfeld_1989, Christensen_Elder_Fogedby_1996, Muratov_2002}.
The potential~$\Psi$ describes long-ranged interactions originating from the reaction-diffusion system.
To study a liquid bridge in this surrogate model we consider two parallel walls embedded in the larger system of volume~$V_\mathrm{sys}$; see \figref{fig:setup}(a).
For simplicity, we only simulate the smaller subsystem between the walls, and assume a constant concentration in the outer system; see \figref{fig:setup}(b).
For each wall separation $s$, we vary $c(\vect r)$ to find the minimal energy $\tilde H_*(s)$ (for details see section I of the Appendix and \cite{Zwicker_2020}).
Consequently, the force~$f$ between the walls is given by $f = -\tilde H_*'(s)$.

\section{Results}

\subsection{Force generated by passive liquid bridge}

We start by analyzing passive liquid bridges, where we expect attractive forces~\cite{fortes_axisymmetric_1982, gouveia_capillary_2022}. 
\figref{fig:result-1-2}(b) shows that this is indeed typically the case, except for very close walls ($s\lesssim 10 w$) where repulsive contact potentials ($g_1<0$, $\vartheta > \frac\pi2$) can lead to overall repulsion ($f>0$).
Our numerical data is consistent with an analytical estimate~\cite{fortes_axisymmetric_1982},
\begin{align}
    f = -\gamma \frac{V\cos(\vartheta)}{s^2} - \gamma \frac{\sqrt{4\pi V}\sin(\vartheta)}{\sqrt{s}}
    \;,
    \label{eq:force-passive}
\end{align}
where $V$ is the conserved volume of the bridge.
Here, the first term dominates for close walls, whereas the second term accounts for the general behavior at larger separation~$s$.
In the simple case of a neutral wall ($\vartheta=\frac\pi2$), the bridge is a cylinder, implying that the free energy is proportional to the surface area, $\tilde H_* \sim \sqrt{sV}$.
The resulting attraction ($f\sim -s^{-1/2}$) thus results from the reduced interfacial area to the dilute region at constant bridge volume~$V$, consistent with \Eqref{eq:force-passive}.

\subsection{Chemical reactions make forces repulsive}

\begin{figure}[t]
    \centering
    \includegraphics[width=\linewidth]{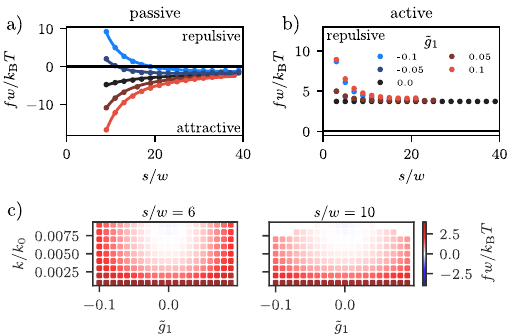}
    \caption{\textbf{Active bridge repel walls.}
(a, b) Force $f$ between walls as a function of separation $s$ for various wall affinities $\tilde{g}_1=g_1 \nu^3/(aw)$ without reactions ($k=0$, panel~a) and with reactions ($k=0.001\,k_0$, panel~b). 
Solid lines in panal a are a fit of \Eqref{eq:force-passive}.
Active bridges break up into separate droplets for large $s$.
(c) $f$ as a function of $\tilde g_1$ and $k$ for $s=6\,w$ and $s=10\,w$.
(a--c) Model parameters are $c_0 V_\mrm{sub} =20000$ (passive) and $c_0=0.1\,\nu^{-1}$ (active), $a=\nu^3 \kBT$, and $k_0=0.001\mobD a \nu^{-2} w^{-2}$.  
}
    \label{fig:result-1-2}
\end{figure}

We next investigate chemically active liquid bridges.
Numerical simulations indicate that bridges are now generally repulsive; see \figref{fig:result-1-2}(b).
Interestingly, the force $f$ is constant for neutral walls ($g_1=0$, $\vartheta=\frac\pi2$), whereas both attractive and repulsive contact potentials increase the repulsion of the walls at small separations~$s$.
The observed forces generally decrease with larger reaction rates $k$; see \figref{fig:result-1-2}(c).
Taken together, chemically active liquid bridge thus exhibit unusual repulsion, which increases with stronger contact potentials of either sign.

\subsection{Increasing repulsion for short wall distances is caused by wall interaction}
To understand the increasing repulsion for small wall separation~$s$, we decompose the total force, $f = f_\mrm{bulk} + f_\mrm{wall} + f_\mrm{react}$, where the bulk and wall forces originate from the volume and surface integrals in \Eqref{eq:free_en_surf}, respectively, and the reactive force is associated with the second term in \Eqref{eq:tot_en_func}.
Analyzing the  individual contributions to the force reveals that the wall term is largest for for small~$s$ (Fig.~5 in Appendix) and thus dominates the overall behavior.
Remarkably, a similar force is observed even without a bridge, indicating that the short-range repulsion is primarily governed by the interaction between the walls and the surrounding dilute phase.
We obtain the associated force~$f_\mrm{wall}^0$ by determining the stationary state profile $c(z)$ without a bridge, and evaluating how the  concentration $c_\mrm{wall}$ at the wall changes with~$s$,
\begin{align}
	\label{eqn:bare_wall_force}
    f_\mrm{wall}^0 
    = 2A_\mathrm{wall} g_1 \partial_s c_\mrm{wall}\;,
\end{align}
where $A_\mathrm{wall}$ is the wall surface area.
To estimate $c_\mrm{wall}$, we approximate the stationary solution of \Eqref{eq:cahn-hilliard-source} by expanding the chemical potential around $c_0$ to linear order in $c$, resulting in the ordinary differential equation (for details see section~III of the Appendix)
\begin{align}
	\label{eqn:bare_wall_ode}
    \partial_z^2 c-\frac{w^2}{4}\partial_z^4 c - \frac{1}{\ell^2}(c-c_0) = 0
    \;,
\end{align}
where the reaction diffusion length scale $\ell=\sqrt{D/k}$ follows from the diffusivity $D=\mobD h''(\cOut^{(0)})=\mobD h''(\cIn^{(0)})$.
Using the boundary conditions $\partial_z c|_{z=\pm s} = \mp g_1/\kappa$ and $\partial_z \mu_{z=\pm s}=0$,
we obtain an analytical solution of \Eqref{eqn:bare_wall_ode}, which allows us to evaluate  $c_\mrm{wall}$ at the walls (section~III in Appendix), and use \Eqref{eqn:bare_wall_force} to determine the associated force $f_\mrm{wall}^0$.
In the limit of weak reactions ($\ell \gg w$), we find 
\begin{multline}
    f^0_\mrm{wall} \approx A_\mrm{wall}\frac{g_1^2}{\kappa}\left[\frac{w^2}{2 s^2}-\frac{2}{\sinh^2(2\frac{s}{w})}\right.
 \\
    \left. +  \left(\frac{\frac{s}{w} \coth\left(2\frac{s}{w}\right)+\frac12}{ \sinh^2(2\frac{s}{w})}-\frac{ w^2}{4 s^2}+\frac{1}{2}\right) \left(\frac{w}{\ell}\right)^2
\right]\;,
    \label{eq:boundary-force}
\end{multline}
which is an approximation up to second order in $w/\ell$, and the full expression is given by Eq.~S27 in the Appendix. 
The force is generally repulsive ($f^0_\mrm{wall}>0$) and decreases with larger wall separation~$s$ since the concentration profile can relax to an energetically more favourable state for larger $s$.
The scaling predicts that the force diverges for small $s$, consistent with \figref{fig:result-1-2}(b).
Moreover, the magnitude $f^0_\mrm{wall}$ depends quadratically on the wall affinity~$g_1$, consistent with the numerical observation in \figref{fig:result-1-2}(c).
In particular, positive $g_1$ implies lower concentration at the wall, whereas negative $g_1$ increases the wall concentration, and both perturbations have similar effects on the free energy and thus the force.
Taken together, we find that wall interactions mediated by the dilute phase explain the repulsive force observed at short separation. 

\subsection{Size control determines capillary forces}
To understand the repulsion at large wall separation~$s$, we next consider a bridge between neutral walls ($g_1=0$, $\vartheta=\frac\pi2$), where the cross-sectional profile is constant.
The total energy is then up to a constant given by 
\begin{multline}
    \tilde H_\mrm{neutral} = s \int \left(h(c)-h(c_0) + \frac{\kappa}{2}|\nabla c|^2\right)\diff A  \\
                       + \frac{ks}{2 \mobD} \int \bigl[c(\vect r) - c_0\bigr]\Psi(\vect r) \, \mathrm{d}A\;,
                       \label{eq:energy_sub}
\end{multline}
where the term $h(c_0)$ in the first line compensates for the outer system (section II in Appendix).
Note that both integrals are over cross-sectional slices, with the first one capturing passive effects whereas the second one accounts for reactions. 
Since the cross-sectional profiles, and thus $\Psi(\vect r)$, do not depend on $s$, the force $f_\mathrm{neutral}=-\diff \tilde H_\mathrm{neutral}/\diff s$ is directly given by the integrals, i.e., the energy required to insert a slice of the liquid bridge between the walls.

\begin{figure}[t]
    \centering
    \includegraphics[width=\linewidth]{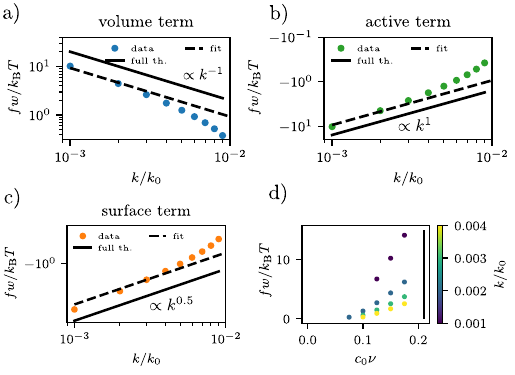}
    \caption{\textbf{Size-control of active bridge causes repulsion.}
    Comparison of individual contributions to the force~$f$ from \Eqref{eq:theory-k-scaling} to numerical data as function of the reaction rate~$k$: the bulk force in~(a), the negative active force in~(b), and the negative surface contribution in~(c).
    (d)~$f$ as a function of the supersaturation~$c_0$ for various~$k$.
    (a--d) Model parameters are $s=5.5\,w$. Other parameters are given in \figref{fig:result-1-2}.
    }
    \label{fig:result-model}
\end{figure}

We estimate the force~$f_\mathrm{neutral}$  using a thin-interface approximation of the bridge~\cite{Weber_Zwicker_2019, Vidal2021}, where we exploit that the bridge radius $R_*$ is governed by size-control~\cite{Zwicker_Suppression_2015}, independent of wall separation~$s$; see \figref{fig:setup}(c).
In particular, the steady-state bridge radius $R_*$ is proportional to the reaction-diffusion length $\ell$, $R_*=\ell \cdot b(c_0, \cIn^{(0)}, \cOut^{(0)})$,
where $b$ is a monotonically increasing function of $c_0$ (section II of the Appendix).
This approach allows us to evaluate all terms of $\tilde H_\mathrm{neutral}$ to obtain
\begin{multline}
    f_\mathrm{neutral} = - 2 \pi \gamma q \sqrt{\frac{D}{k}}  +\Delta h \pi b^2 \frac{D}{k}
\\
      - \frac{\pi b^4 D^2 (c_\mathrm{in}^\mathrm{(0)}-c_0)^2}{4 \mobD k}\left[
      	 \log \left(\frac{2}{b}\right)-\gamma_\mathrm{eul} -\frac{1}{4}
	\right]\;,
      \label{eq:theory-k-scaling}
\end{multline}
with $\Delta h = h(c_0)-h(\cIn) + h'(c_0)(\cIn - c_0)$ and the Euler--Mascheroni constant $\gamma_\mathrm{eul}\approx 0.5772$ (sections II-III of the Appendix).
The first term in \Eqref{eq:theory-k-scaling} corresponds to the force arising from minimization of interfacial energies, the second term captures the gain from the increased bridge volume, and the third term accounts for the contribution of chemical reactions.

Our theory predicts that the repulsive force~$f_\mathrm{neutral}$ exerted by the bridge decreases with larger reaction rates~$k$, consistent with numerical simulations at fixed wall separation; see \figref{fig:result-model}(a--c).
In particular, the separate terms in \Eqref{eq:theory-k-scaling} predict the respective contributions of bulk, surface, and reactions at small $k$, but they overestimate $f_\mathrm{neutral}$ for large $k$.
The overall trend can be understood qualitatively:
Repulsive forces originate from the free energy gain of elongating the bridge since phase separation is favorable. 
Moreover, larger $k$ implies narrower bridges, leading to smaller free energy gains and weaker forces.
This qualitative picture predicts larger forces for stronger supersaturation, i.e., larger background concentration $c_0$.
\figref{fig:result-model}(d) indeed shows a monotonic increase in the repulsive force with increasing $c_0$.

Taken together, the force exerted by a chemically active  bridge can be understood as a superposition of two contributions, $f = f_\mathrm{neutral} + f^0_\mrm{wall}$.
Size-control implies a repulsion associated with the bridge ($f_\mathrm{neutral} > 0$), which is independent of wall separation since it is energetically favorable to elongate the bridge; see \Eqref{eq:theory-k-scaling}.
Additionally, the direct interactions with the walls modify the field in their vicinity, accruing an additional energy penalty leading to the repulsion described by \Eqref{eq:boundary-force}, $f^0_\mrm{wall}>0$.
Combining these effects leads to an overall repulsive force that decreases with separation~$s$ (\figref{fig:result-1-2}b).

\section{Conclusion}
In summary, we demonstrated that active chemical reactions alter the forces exerted by a liquid bridge profoundly:
While passive bridges typically attract the structures they connect, active bridges repel them.
Interestingly, the repulsive forces of active bridges are largely constant, except for separations comparable to a few interfacial widths.
Intuitively, the general repulsion results from size-control, which limits the amount of material that can segregate into the liquid bridge.
Larger separations allow more material to segregate, lowering the energy of the overall supersaturated environment, which leads to repulsion.

We studied the simple situation of a single active bridge connecting two rigid, flat walls.
However, cellular boundaries are often curved and deformable.
Moreover, we focused on the quasi-stationary limit, but real dynamics can be fast, where capillary-driven forces and size-controlled forces might compete. 
While addressing all these features requires advanced computational methods, we expect our main findings are robust.
In particular, the chemically active liquid bridges we studied provide a first example of force generation in non-equilibrium phase-separation. 
While other non-equilibrium phase-separating systems, such as active Model B+~\cite{Cates_Nardini_2023}, will likely also show modified forces, the relation between these active field theories remains to be understood.
In any case, active liquid bridges provide an exciting mechanism to push apart structures, which might be employed by cells to control organelle placement and morphology.

\begin{acknowledgments}
We thank Yicheng Qiang for insightful discussion and Lukas Kristensen for critical reading of the manuscript.
We gratefully acknowledge funding from the Max Planck Society, the German Research Foundation (DFG) under grant agreement ZW 222/3, and the European Research Council (ERC, EmulSim, 101044662).
\end{acknowledgments}

\appendix
\onecolumngrid

\section{Numerical force calculation}
Here, we describe the numerical scheme used to calculate the forces a liquid bridge exerts on two walls. 
To study the liquid bridge forces, we consider a system with two movable walls embedded in a larger volume, as shown in Fig.~1(a) of the main text.
The volume $V_\mrm{sub}$ of the subsystem can be changed by adjusting the separation~$s$ between the two wall segments.
The average concentration is given as $c_0$.
Assuming an infinitesimal slow movement of the walls, such that the bridge can equilibrate, we can define a generalized force $f$ from the energy difference of the equilibrium states at different separations,
\begin{align}
    f &= -\frac{\min_{c, s=s_2} H[c] -  \min_{c, s=s_1}H[c]}{s_2-s_1}
     = -\frac{\partial H}{\partial s}|_{c_\mathrm{min}}\;,
\end{align}
where $H$ is the free energy of the full system and $c$ is the concentration of the bridge-forming component.  
We simulate the subsystem for different separations $s_1$ and $s_2$ with an initialized bridge and equilibrate the system by solving Eq.~4 of the main text using py-pde \cite{Zwicker_2020}.
We then compute the generalized force from the equation above. 
The exact setup of the system differs between passive and active systems, which will be explained in the following sections.

\subsection{Force calculation in passive system}
In the passive case, the liquid bridge concentration is set by the equilibrium values $\cInN$ inside and $\cOutN$ outside. 
We only need to simulate the subsystem between the two walls given by $V_\mrm{sub}$ as the concentration outside is constant and given by the equilibrium concentration $\cOutN$.
The free energy of the full system can be written as 
\begin{align}
    H = \int_{V_\mrm{sub}} \left(h(c) + \frac{\kappa}{2}|\nabla c|^2 \right) \diff V + (V_\mrm{sys}-V_\mrm{sub})h(\cOutN)\;.
\end{align}
We generally analyze an overall system with average composition $c_0$, but since we only simulate the subsystem without explicitly simulating the outer part, it is necessary to adjust the average concentration $\bar c_\mathrm{sub}$ according to the wall separation. 
This can be seen by considering the total amount of material,
\begin{subequations}
\begin{align}
    \Vsys c_0 &= \int_{\Vsub}c_\mrm{sub}(\boldsymbol{r}) \diff V + (\Vsys -\Vsub)\cOutN\;,\\
    \Vsys (c_0 - \cOutN) &= \int_{\Vsub} (\cSub(\boldsymbol{r}) -\cOutN) \diff V= (\bar{c}_{\mrm{sub}}-\cOutN)\Vsub\;,
\end{align}
\end{subequations}
so that the average concentration inside the sub-volume must be
\begin{align}
    \bar{c}_\mrm{sub}=\frac{\Vsys}{\Vsub}\left(c_0 - \cOutN\right) + \cOutN\;.
\end{align}
For the simple free energy we chose in Eq.~(2) in the main text, this expression simplifies to $\bar{c}_\mrm{sub}\Vsub=\Vsys c_0$ so that the mean concentration in the subsystem is set by the total mass of bridge material in the system.
In this case, the compensation term for the outside vanishes.

\subsection{Force calculation in active system}
For the active system, we assume that the large system outside the subsystem remains homogeneous; therefore, we do not need to simulate it explicitly.
In the active case, the concentration outside the subsystem is given by $c_0$ as the chemical reactions dominate away from the bridge and set the concentration.
Consequently, the energy of the full system with the compensation term is given by
\begin{subequations}
    \begin{align}
        H &= \int_{V_\mathrm{sub}} \left( h(c) +\frac{\kappa}{2}|\nabla c|^2\right) \mathrm{d}V + (V_\mathrm{sys}-V_\mathrm{sub})h(c_0)\;,\\
        &= \int_{V_\mathrm{sub}} \left(\tilde{h}(c) +\frac{\kappa}{2}|\nabla c|^2\right) \mathrm{d}V + \text{const}\;,
    \end{align}
\end{subequations}
with $\tilde{h}(c) = h(c)-h(c_0)$.
Here, it is not necessary to adapt the mean concentration inside the subvolume, as the mean concentration of the outside system is given by $c_0$, and consequently, the mean concentration inside the subvolume is also given by $\bar{c}_\mrm{sub}=c_0$.
For linear reactions, we can employ a mapping of the active system to an equivalent equilibrium system by defining 
an augmented free energy functional
\begin{align}
       \tilde H[c] = H[c]  + H_\mathrm{react}[c]
       \;,
\end{align}
where
\begin{align}
    H_\mathrm{react}[c] &= \frac{k}{2 \mobD} \int \bigl[c(\vect r) - c_0\bigr]\Psi(\vect r) \, \mathrm{d}\vect{r}
       \label{eq:energy_react}
\end{align}
captures the energy associated with reactions~\cite{Liu_Goldenfeld_1989, Christensen_Elder_Fogedby_1996, Muratov_2002}. 
Here, $\Psi$ is the solution to the Poisson equation $\nabla^2 \Psi = c_0-c(\vect r)$ and describes the long-ranged interactions, which originate from the interplay of chemical reactions and diffusion in the original model.
Since this surrogate model requires mass conservation, we solve for $\Psi$ employing Neumann boundary conditions, $\vect{n}.\nabla \Psi=0$.

\section{Force calculation using thin interface approximation}

In the following, we present a derivation of the force exerted by a liquid bridge with neutral wall interactions, which implies a straight cylinder geometry. 
We will solve for the force within the thin interface approximation following the Supporting Information of reference \cite{Vidal2021} and \cite{Ziethen_Kirschbaum_Zwicker_2023} 
Since the force $f$ is defined as derivative of the free energy, $f=- \partial_s H$,  we will first derive an expression for the free energy $H$ of a liquid bridge only depending on the bridge radius $R$ and separation $s$. 

 \subsection{Free energy in passive system}
 We consider a liquid bridge forming between two walls separated by $s$, which form a subsystem of volume $V_\mathrm{sub}$. 
The full system has the volume $\Vsys$ and the average concentration in the entire system is given by $c_0$.
We assume that a thin interface separates the bridge and the surrounding bulk phase, and that the concentration inside the bridge is given by $\cIn$ and the concentration outside the bridge is given by $\cOut$. 
 In the passive case, the concentration inside and outside of the bridge are simply given by the equilibrium concentrations $\cIn^{(0)}$ and $\cOut^{(0)}$ in the simple case where surface tension effects are negligible. 
 The bridge volume is determined by the total amount of material,
 \begin{align}
    V = \frac{\Vsys(c_0 - \cOut^{(0)})}{\cIn^{(0)}-\cOut^{(0)}}\;. 
 \end{align}
We can then write 
\begin{align}
    H\approx - V \left(h(\cOut^{(0)})-h(\cIn^{(0)})\right) + \gamma A\;,
\end{align}
where $h$ denotes the free energy density and we express the surface energy between the two phases via the surface tension $\gamma$ and the contact area $A$ between the two phases. 

\subsection{Local contributions to free energy in active system}
In the active system, the situation is different as the chemical reactions force the system to have a concentration $c_0$ far away from the bridge. 
Inside the bridge, the concentration gradually increases in radial direction from a value below $\cIn^{(0)}$ at the centerline to the dense phase equilibrium concentration $\cIn^{(0)}$ at the interface.
In contrast, outside the bridge, the concentration field exhibits a gradient that transitions from the equilibrium concentration $\cOut^{(0)}$ at the bridge interface to the mean concentration $c_0$ far away from the bridge~\cite{Zwicker_Paulin_Burg_2025}.
Within the individual regions, the concentration varies slowly on the order of the reaction diffusion length scale~$\ell$, which we assume to be large compared to the bridge radius~$R$.
We thus approximate the concentration profiles by constant values to evaluate the integral given in \Eqref{eq:tot_en_func}.
In particular, we approximate the concentration inside the bridge by the equilibrium concentration $\cIn^{(0)}$, and the concentration of the outer system by $c_0$.
In contrast, the concentration outside the bridge in the subsystem is given by $\cOut$, which will be determined such that the average concentration remains $c_0$, implying
\begin{align}
    (\Vsub - V)\cOut + V \cIn^{(0)} = \Vsub c_0\;.
\end{align}
We thus find
\begin{align}
    \cOut = \frac{\Vsub c_0 - V \cIn^{(0)}}{\Vsub-V}\;.
\end{align}
We can thus write the total free energy of the system as 
\begin{align}
    H\approx V h(\cIn^{(0)})+(\Vsub-V)h(\cOut) + (\Vsys-\Vsub)h(c_0)+\gamma A\;,
\end{align}
where $h$ denotes the free energy density, and we express the surface energy between the two phases via the surface tension $\gamma$ and the contact area $A$ between the two phases. 

We assume the bridge to be much smaller than the surrounding phase, so the concentration $\cOut$ in the dilute phase will be close to the mean concentration $c_0$.
We thus expand  the free energy density of the dilute phase as
 \begin{align}
     h(\cOut) \approx h(c_0) + \left.\frac{\partial h}{\partial c}\right|_{c=c_0}(\cOut-c_0)
     \;.
 \end{align}
 Inserting these expressions into the full free energy, we arrive at
 \begin{align}
     H &\approx V h(\cIn^{(0)}) + (\Vsub-V)\left[h(c_0) + \left.\frac{\partial h}{\partial c}\right|_{c=c_0}\biggl(\frac{\Vsub c_0 - V \cIn^{(0)}}{\Vsub-V}-c_0\biggr)\right] + (\Vsys-\Vsub)h(c_0)+\gamma A\;,
\notag\\
       &\approx V \left(h(\cIn^{(0)})-h(c_0) + h'(c_0)(c_0 - \cIn^{(0)})\right) + \gamma A + \text{const}
\notag\\
        &\approx -\Delta h V + \gamma A\;,
 \end{align}
 where $\Delta h = h(c_0)-h(\cInN) + h'(c_0)(\cInN - c_0)$.

\subsection{Non-local contributions to free energy in active system}
In the treatment above, we have only considered the local energy contributions.
However, we also need to consider the additional energy contribution of the chemical reactions described by the surrogate model.
To do so, we employ the cylindrical symmetry of the system to write
\begin{align}
    H_\mathrm{react} = & s \left(\frac{k}{2 \mobD} \int \bigl[c(\vect r) - c_0\bigr]\Psi(\vect r) \, \mathrm{d}A \right)\;.
    \label{eq:h-react}
\end{align}
In the sections above, we have solved the local terms of the free energy in the thin interface approximation, assuming constant concentrations inside and outside the droplet. 
This approximation is not compatible with the reactive energy and we thus use a more refined approximation to estimate this last contribution.
We do this by approximating the concentration profiles inside and outside using an effective droplet model.
Within this model we can solve for the concentration profile in the dilute and dense phase separately by approximating the chemical potential as $\mu \approx \mu(c_i^{(0)}) + \partial_c \mu|_{c_i^{(0)}}(c_i^{(0)})(c-c_i^{(0)})-\kappa \nabla^2 c$, where $i=\mrm{in}/ \mrm{out}$. 
We then get the following expression for the steady state solution of the concentration field for the inside and outside phase
\begin{align}
    \nabla_\mathrm{2d}^2 c - \frac{1}{\ell^2}(c-c_0) = 0\;,
\end{align}
where $\nabla_\mathrm{2d}$ is the nabla operator in the 2d cross-section, and $\ell=\sqrt{D/k}$ denotes the reaction diffusion length scale, where the diffusivity is given by $D=\mobD f''(\cOutN)=\mobD f''(\cInN)$.
We can then obtain the steady state solutions of the concentrations with  boundary conditions $\cIn'(0)=0$, $\cIn(R)=\cInN$, $\cOut(R)=\cOutN$, and $\cOut(r \to \infty)=c_0$,
\begin{subequations}
\begin{align}
    \cIn(r) &= c_0 + \frac{(\cIn^{(0)}-c_0) I_0(r/\ell)}{I_0(R/\ell)}\;, \label{eq:cin}\\
    \cOut(r) &= c_0 + \frac{(\cOut^{(0)}-c_0) K_0(r/\ell)}{K_0(R/\ell)}\;. \label{eq:cout}
\end{align}
\end{subequations}
To calculate the reactive part of the free energy, we employ the concentration profiles above and insert them into \Eqref{eq:h-react}, which gives \cite{Ziethen_Zwicker_2024}
\begin{align}
    H_\mathrm{react}=s \frac{\pi k(c_\mathrm{in}^\mathrm{(0)}-c_0)^2}{8 \Lambda_\mathrm{d}}\left({2 \log \left(\frac{2 \ell}{R}\right)-(2 \gamma_\mathrm{eul} +\frac{1}{2})}\right)R^4=s f_\mrm{react}\;,
\end{align}
where $\gamma_\mathrm{eul}\approx 0.577$ is the Euler--Mascheroni constant.

\subsection{Total force in active system}
\begin{figure}[t]
    \centering
    \includegraphics[width=0.7\textwidth]{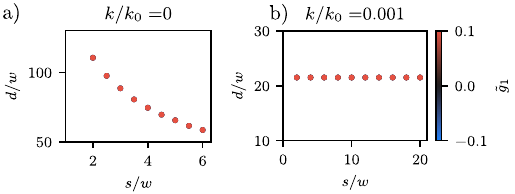}
    \caption{\textbf{Chemical reactions imply constant bridge diameter.} (a, b) Diameter $d$ of the bridge as a function of wall separation~$s$ for passive system (panel a) and active system (panel b) for various wall affinities $\tilde g_1$.
     The total mass for the passive system is given by $V_\mrm{sub}c_0 = 10000$, other model parameters are given in Fig. 2 of the main text.} 
    \label{fig:si-diams}
\end{figure}
We can now write the total force of a bridge with bridge radius $R$ as a sum over all contributions,
\begin{align}
     f = -\frac{\diff \tilde H}{\diff s} &\approx \Delta h A -\gamma U
     -f_\mathrm{react}\;,
\end{align}
where $A=2\pi R^2$ is the cross-sectional area of the bridge and $U=2\pi R$ denotes its circumference.
In steady state, the bridge radius $R_*$ is given by a balance of phase separation and chemical reactions.
The steady state radius of a chemically active bridge can be estimated using the effective model, which results in~\cite{Zwicker_Suppression_2015}
\begin{align}
    R_* &=\ell \, b \left(\frac{c_0-\cOutN}{\cInN-c_0}\right)
   && \text{with} &
    b^{-1}(x)=\frac{I_1(x)K_0(x)}{I_0(x)K_1(x)}
\end{align}
where the function $b$ is defined by its inverse given on the right.
We also checked whether this theoretical prediction can be observed in our simulation, and indeed, we find a controlled central bridge diameter independent of the separation $s$ in the presence of active chemical reactions, whereas we observe a decreasing bridge diameter in the passive case; see \figref{fig:si-diams}. 
Using the constant bridge radius, we write the total force as
\begin{align}
    f_\mrm{neutral}  &\approx -2\pi\Delta h b^2 Dk^{-1} + 2 \pi \gamma b \sqrt{D/k} 
     -f_\mathrm{react}\;,
\end{align}
where we omit the argument of the function $b(x)$ for brevity.
Here, 
\begin{align}
f_\mathrm{react} = \frac{\pi k(c_\mathrm{in}^\mathrm{(0)}-c_0)^2}{8 \Lambda_\mathrm{d}}\left({2 \log \left(\frac{2}{b}\right)-(2 \gamma_\mathrm{eul.} +\frac{1}{2})}\right)b^4 \left(\frac{D}{k}\right)^2
\end{align}
This expression contains the diffusivity~$D$ and the diffusive mobility~$\mobD$, which are not independent.
We choose to replace $D$ by the more microscopic parameters of the original equations of motion $D=\mobD f''(\cOutN)=\mobD a$ to obtain
\begin{align}
    f_\mathrm{neutral} = - 2 \pi \gamma b \sqrt{\frac{D}{k}}  + \Delta h \pi b^2 \frac{D}{k}
      - \frac{\pi b^4 D^2 (c_\mathrm{in}^\mathrm{(0)}-c_0)^2}{4 \mobD k}\left[
        \log \left(\frac{2}{b}\right)-\gamma_\mathrm{eul} -\frac{1}{4}
 \right]\;,
\end{align}
which is equivalent to \Eqref{eq:theory-k-scaling} in the main text.

\section{Analytical calculation for wall interaction}

\begin{figure}[t]
    \centering
    \includegraphics[width=0.7\textwidth]{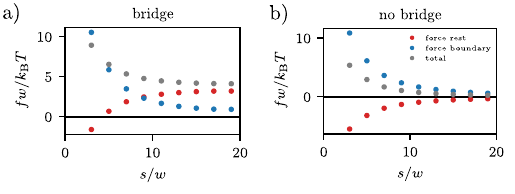}
    \caption{\textbf{Wall interaction dominates short range repulsion.} (a, b) Individual force contributions as a function of wall separation $s$ with a bridge (panel a) and without a bridge (panel b).
    The total force is shown in gray, the boundary term (second term of Eq.~1 in the main text) in blue, and the sum over all other contributions in red.
    For both systems the boundary term dominates the small distance behavior of the total force. The wall affinity is set to $\tilde g_1=0.1$, other model parameters are given in Fig.~2 of the main text.}
    \label{fig:s1}
\end{figure}

Fig.~2(b) of the main text shows that the repulsive force in the presence of active chemical reactions diverges for small wall separations. 
\figref{fig:s1} shows that walls connected by a dilute phase also exhibit repulsive forces showing the same behavior at small separations as in the case of a bridge. 
The figure also reveals that the largest contribution stems from the boundary term.
To investigate this effect further, we study walls connected by a dilute phase (without a bridge) in detail.
We can solve for the concentration profile in the dilute phase by approximating the chemical potential as $\mu \approx \mu(\cOut^{(0)}) + \mu'(\cOut^{(0)})(c-\cOut^{(0)})-\kappa \nabla^2 c$. 
We then get the following expression for the steady state solution of the concentration field,
\begin{align}
    \partial_z^2 c-L_1^2\partial_z^4 c - \frac{1}{\ell^2}(c-c_0) = 0\;,
\end{align}
with  boundary conditions $\partial_z c|_{z=\pm s} = \mp g_1/\kappa$ and $\partial_z \mu_{z=\pm s}=0$.
The length scale is defined as $L_1=\sqrt{\kappa/h''(c_\mrm{eq})}$ and coincides with half of the interface width for our choice of free energy, which will be used in the following.
The second length scale is the reaction diffusion length scale $\ell=\sqrt{D/k}$.
For these conditions,
\begin{align}
    \cOut = c_0 + \frac{g_1}{\kappa(\Gamma_1^2-\Gamma_2^2)}\left( \frac{\cosh(\Gamma_1 z)}{\sinh(\Gamma_1 s)} \Gamma_1 - \frac{\cosh(\Gamma_2 z)}{\sinh(\Gamma_2 s)} \Gamma_2\right)\;,
\end{align}
with 
\begin{align}
    \Gamma_{1/2}=\frac{\sqrt{2}}{w} \sqrt{1 \pm \sqrt{1-\left(\frac{w}{\ell}\right)^2}}\;.
\end{align}
We can now estimate the forces resulting from such a concentration profile.
The dominant contribution in the numerical simulations stem from the boundary term.
The boundary force can be simply estimated as 
\begin{align}
    f_\mrm{wall}^0 = 2 A_\mrm{wall} g_1 \partial_s c_\mrm{int}\;,
\end{align}
where $c_\mrm{int}$ is given by the concentration evaluated at $z=\pm s$ and $A_\mrm{wall}$ is the surface area of the wall.
Hence,
\begin{align}
    f_\mrm{wall}^0 = 2 \frac{g_1^2}{\kappa(\Gamma_1^2+\Gamma_2^2)}\left(\frac{\Gamma_1}{\sinh^2(\Gamma_1 s)} - \frac{\Gamma_2}{\sinh^2(\Gamma_2 s)} \right)\;.
\end{align}
We can expand this expression for small $w/\ell$ to arrive at Eq.~(9) of the main text.

Note that this repulsion stems from the interplay between the boundary condition minimizing the surface energy and the resulting reactive fluxes, which cannot vanish on a local level.
Instead, there is a spatially varying concentration field at low concentrations where diffusive and reactive fluxes vanish. 
This concentration profile can relax to an energetically more favourable state if the wall distance is increased resulting in repulsive forces.
The magnitude of these repulsive forces depends quadratically on the wall affinity, which we also find in our numerical force calculations; see Fig.~2(c).

\bibliographystyle{apsrev4-1}
\bibliography{lit_v2}

\end{document}